# Defense Against Flooding Attacks using Probabilistic Thresholds in the Internet of Things Ecosystem

Seyed Meysam Zarei [1] . Reza Fotohi[2]

[1]Department of Computer, Isfahan (Khorasgan) Branch, Islamic Azad University, Isfahan 81595-158, Iran

[2]Faculty of Computer Science and Engineering, Shahid Beheshti University, G. C. Evin, Tehran 1983969411, Iran

**Correspondence**
Faculty of Computer Science and Engineering, Shahid Beheshti University, G. C. Evin, Tehran 1983969411, Iran

Email: Fotohi.reza@gmail.com;
R_fotohi@sbu.ac.ir

**Abstract** The Internet of Things (IoT) ecosystem allows communication between billions of devices worldwide that are collecting data autonomously. The vast amount of data generated by these devices must be controlled totally securely. The centralized solutions are not capable of responding to these concerns due to security challenges problems. Thus, the Average Packet Transmission RREQ (APT-RREQ) as an effective solution, has been employed to overcome these concerns to allow for entirely secure communication between devices. In this paper, an approach called LSFA-IoT is proposed that protects the AODV routing protocol as well as the IoT network against flooding. The proposed method is divided into two main phases; The first phase includes a physical layer intrusion and attack detection system used to detect attacks, and the second phase involves detecting incorrect events through APT-RREQ messages. The simulation results indicated the superiority of the proposed method in terms of False Positive Rate (FPR), False Negative Rate (FPR), Detection Rate (DR), and Packet Delivery Rate (PDR) compared to REATO and IRAD. Also, the simulation results show how the proposed approach can significantly increase the security of each thing and network security.

**Keywords** Internet of Things (IoT) . Flooding attacks . Routing security . Average Packet Transmission

## 1 Introduction

The succeeding generation of internet services present everywhere and influencing all aspects of our diary life will be developed by numerous reasons including improvements in social network technologies, mobile and extending computing, and exponential evolution in Internet facilities [1, 2]. It can be expected that by 2031the number of IoT devices will surpass over 51 billion. However, despite the advantages of IoT through different applications, they are potentially vulnerable against risks as a result of circumstances, in which the actions are monitored by no pilot. It intensifies the necessity of designing reliable and secure IoT as well as overcoming the challenges to prevent mutilation and destruction to the other systems and human lives [3, 4]. Some attacks like flooding attack (FA) enter the system illegally. By affecting attack in an IoT ecosystem, it is difficult to remove the threat and make the system online again. It is worth to state that the usual approaches to secure information including intrusion detection or encryption, are not enough for coping with such risks. Besides, problem of previous IoT systems was the mere attempt on eliminating a single attack type and were only resistant to it. If the system was subject to combined attack, it would be practically inoperative, and the intrusion operation would fail the system quickly.

In LSFA-IoT, we present a new security system that avoids producing unnecessary *RREQ* packets by malicious nodes. Our system can identify the attacks sending a large number of fake *RREQ* packets with invalid IPs to the destination. To detect these malicious nodes, we propose to implement a security module in each certified node in the network. This solution consists of two basic phases: the first phase detects and identifies loading attacks, and in the second phase we eliminate the attacks discovered in the first phase so that they do not carry out malicious activity on the network. Figure 1 shows a vulnerable IoT connected scenario.

**Fig. 1** A vulnerable IoT connected environment [5].

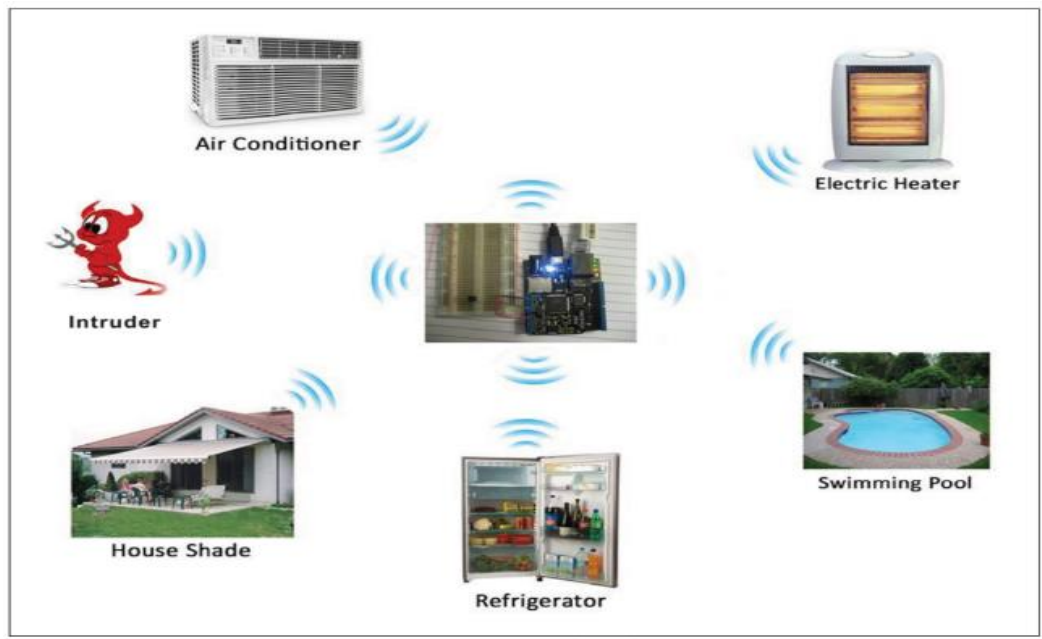

The rest of this research is as follows: In Section 2, fundamental concepts, which are used in this paper, are presented. Section 3 describes the proposed method in full detail. The simulation results are given in Section 4 of this paper, and finally, the conclusions and future work of this article are discussed in Section 5.

## 2  Relevant terms

In this section, the basic concepts of application scenarios, security threats targeting IoT, and detection schemes are explained.

## 2.1 Application Scenarios

There are a variety of areas composing of industry, municipal substructure, smart surroundings, and healthcare field for the application scenarios of the IoT (c.f. Fig. 2) [6, 7]. These scenarios suffer the attacks that are various, cross-cutting across lots of procedures layers in IoT structural design, and containing incorporation of a diversity of attack techniques which will lead to increase the analysing intricacy of the IoT security. In addition, the incentive of the attacker is probably different in various application scenarios, for instance, the target might be gain entrance to critical user data in a wearable application, despite the fact that healthcare-related attacks want to deteriorate the life safety of patients.

**Fig. 2** Description of application scenarios in the IoT [6].

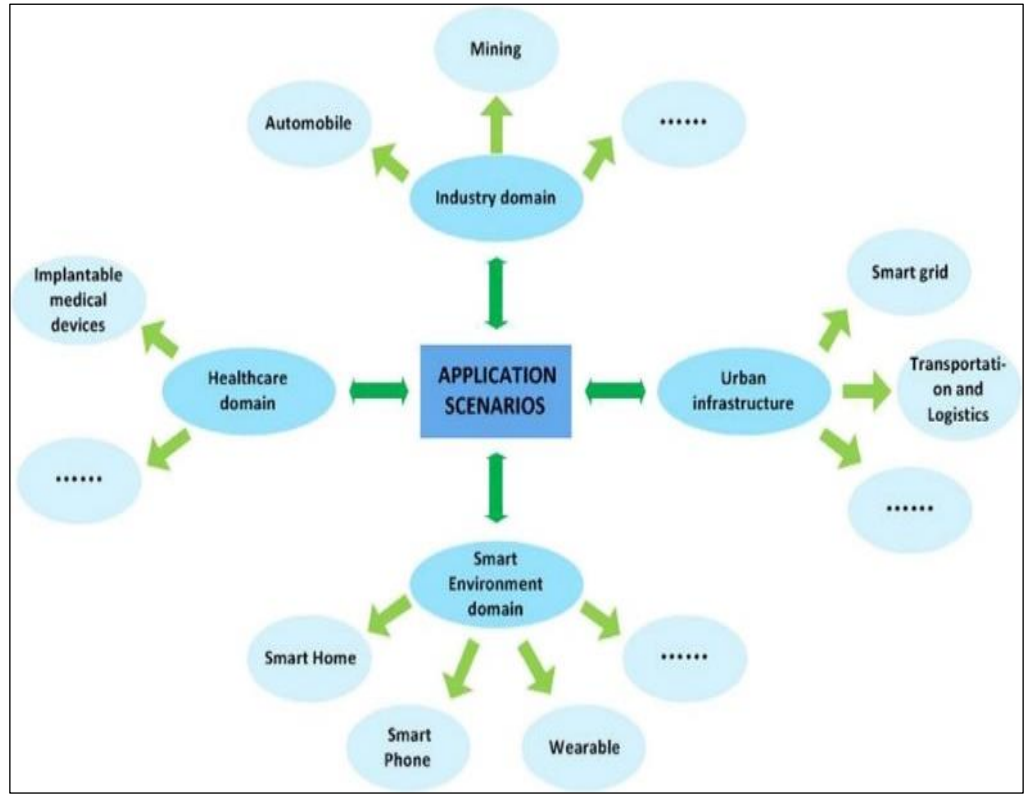

## 2.2 Security Attacks

In this article, the following attacks are discussed:

- *Flooding Attack* plays a key role in the security of IoT so that it is facile to begin; however, challenging to stop. A mischievous node can launch an attack with simplicity using sending an immoderately high number of route request (*RREQ*) packs or inoperable data ones to unreal destinations. Therefore, the network is rendered impractical so that all its resources are undermined to serve this hurricane of *RREQ* packs; as a result, it cannot carry out its standard routing responsibility [6, 7]. The flooding attacks are shown in Figure 3.

**Fig. 3** The flooding attacks in the IoT [7].

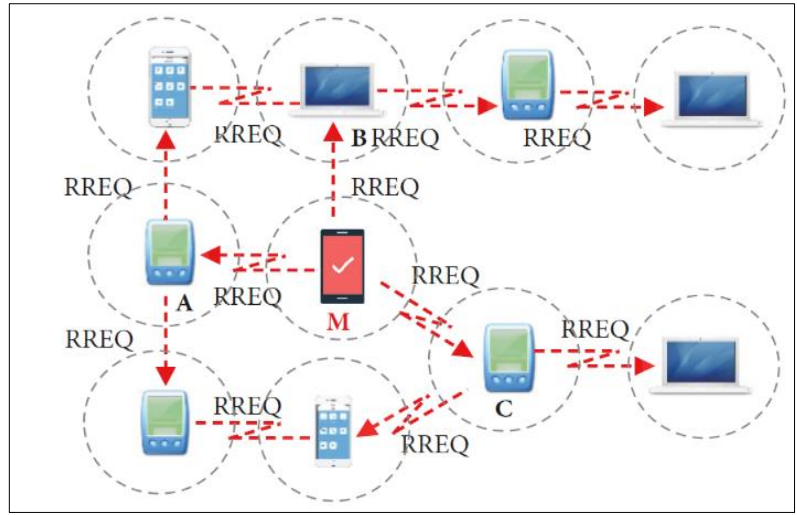

## 2.3 Detection schemes

A large number of studies have been conducted on securing communication between IoT devices. According to the subject matter of the present research, several prominent studies performed in this domain will be reviewed and evaluated.

A method to protect the Internet of Things against cyber-attacks has been developed by Pacheco et al. [8]. Initially, they divided the IoT security structure into four-tier SIs, including devices (end nodes), networks, services, and applications. At the time, the authors' proposed method was developed to create a general threat model for identifying weaknesses at each layer and potential countermeasures that could be extended to extend the benefits of using them.

In this study, a set of practices and attitudes in IoT security have been evaluated in [9]. IoT security allows us to know about a number of tests that are no different from guaranteeing the security of other computing devices such as laptops, servers, or even mobile devices. In particular, they develop two categories of security attacks based on the IoT system. The first is the attack on the four-tier structural design of the Internet of Things. The second case of IoT sensitivity and security depends on various scenarios, including the systematic basis for protecting different applications.

With the aim of implementing network-based encryption, the Ring-LWE scheme, which was developed as a new, network-based encryption in [10] to replace public encryption, was proposed by the authors. However, malicious side channel attacks can terribly damage Ring-LWE performance. In this study, using 16-bit microcontrollers, attack scenarios were analyzed and using bit analysis for the IoT ecosystem, a countermeasure was proposed. There must be an optimization to use this design in IoT devices via 8-bit, 32-bit or 64-bit microcontrollers.

In [11] the authors proposed a security method to validate and connect the main generation between IoT devices. They have comprehensively created a master plan for generating and identifying the physical layer, which is based on frequency jump communications because RSS generates specific frequencies of its own set of parameters. In addition, they considered the feasibility of actual executions to examine and evaluate the proposed method.

The authors discuss in detail the history of the Internet of Things, as well as the security-based analysis of the Internet of Things ecosystem [12]. They provide two types of security classifications in the IoT ecosystem and different protection mechanisms that can provide a better understanding of the issue, the current clear space, and future research guidelines for protecting the IoT from various attacks. The proposed method has the best performance in terms of operational power, False Positive Rate and False Negative Rate.

In [13] devastating attacks on the IoT ecosystem and how IoT devices are moving have added to the difficulty of detecting such malicious attacks. The present paper provides a model for maintaining the path of IoT devices and identifying destructive nodes on the Fog infrastructure. In the proposed method, a Fog computing layer is provided to monitor and detect malicious attacks in IoT environments.

In [14], they suggest a way to dynamically detect DoS attacks in the IoT ecosystem. Since this article needed to find a solution to defend and secure the IoT ecosystem against DoS attacks, considering all the possible conditions for malicious attacks that may occur, they offered a safe solution to defend against DoS attacks.

In [15], the authors suggest a deep learning-based approach that can detect routing protocol attacks based on large-scale, high-distribution data. Flooding attack is successfully identified by the proposed method and is excluded from routing operations.

In [16], the solution designed to fit the NPS architecture is authenticated using a real-world test platform and built by an NPS prototype that receives open data in real time through a set of compatible sources. Slowly This method, called REATO, is used to detect and thwart a DoS attack compared to the IoT middleware known as NPS. Work has begun on the need to discover a solution to protect an IoT system from DoS attacks, taking into account all potential contingencies that may occur.

A machine learning technique with deep learning is presented in [17], which detects routing attacks. A deep, high-scalability learning-based approach is recommended to detect IoT routing attacks that, with astonishing accuracy and precision, reduce degree attacks, flood greetings, and number of copies. Using Cooja IoT emulator, high-fidelity attack data is generated in IoT networks containing 10-1000 nodes. To use in-depth learning for IoT in cybersecurity, the availability of significant IoT attack information is essential.

Table 1 lists the type of attacks, and placement schema of previous research.

**Table 1** Previous research

| Ref. | Placement schema | Attack |
| --- | --- | --- |
| [3] | Centralized | DoS |
| [4] | Centralized | Routing attack |
| [5] | Distributed | Side-channel attack |
| [6] | Hybrid | Physical-layer attack |
| [7] | Hybrid | Multiple conventional attacks |
| [8] | Distributed | Impersonation and replay attacks |
| [9] | Hybrid | Flooding attack |
| [10] | Centralized | Flooding attack |
| [11] | – | DDoS |

# 3 The proposed LSFA-IoT schema

The proposed method has six steps as follows: Section 3.1 lists hypotheses of the proposed method. Overview of the LSFA-IoT schema is discussed in Section 3.2. Section 3.3 describes the Misbehavior notification step in detail. The mechanism to detect a physical-layer attack in LSFA-IoT is given in Section 3.4. Section 3.5, Adding the malicious thing to detention list is described in detail. And in the last section, Revision of malicious thing is discussed.

## 3.1 Hypotheses of the proposed method
The parameters that we have considered for our proposed method are as follows.

- There isn't any central controller in the IoT network.
- All the things in the IoT network act as a final system and router for sending packets.
- All the things in the IoT network are mobile.
- The connections between the things are done by the AODV [24] protocol.
- Things should follow the standard protocol to join or left.

## 3.2 Overview of the LSFA-IoT schema
The proposed LSFA-IoT method is designed for the IoT based on the AODV protocol. LSFA-IoT is based on the analysis and prevention of the flooding attack in the network layer in IoT. The LSFA-IoT is based on the neighbor suppression technique which detects the malicious thing during the route building step. In the case of finding a malicious thing, the proposed method keeps that solitary for a while and checks its behavior to avoid flooding attack in the network layer. Before sending a packet, each node checks the detention list field in LSFA-IoT. If the destructive node is in the arrest list, the data packet will not be forwarded to it, otherwise it will be forwarded depending on it.

In LSFA-IoT, we present a new security system that avoids producing unnecessary *RREQ* packets by malicious nodes. Our system can identify the attacks sending a large number of fake *RREQ* packets with invalid IPs to the destination. To detect these malicious nodes, we propose to implement a security module in each certified node in the network. This solution consists of two basic phases: the first phase detects and identifies loading attacks, and in the second phase we eliminate the attacks discovered in the first phase so that they do not carry out malicious activity on the network.

The first part is used to stabilize the status of the network. If the number of route requests exceeds the threshold, we will make these nodes aware of misbehavior and abnormal behavior in the network. The announcement indicates that there is a flooding attack in the IoT ecosystem and will lead to the start of the second phase.

The duty of the second part is to discover the misbehavior sources in the network that can be a single attack or common torrent attacks. Such attacks can be detected based on the immediate routings of the different packets sent/received by each existing node. The AODV protocol uses a voluntary *Hello* message for the stability of the connections between the neighbor nodes. We use the

Hello packet to send information like starting route discovery by network nodes. All the nodes must observe all the mechanism defined to avoid creating fake route requests in the network.

### 3.3 Misbehavior notification step

The misbehavior notification step is used to optimally detect the misbehavior of the nodes and contribute to creating active security solutions. The detection system is inactive as long as the network is in secure status and no flooding attack is reported. In order to clarify the status of the IoT ecosystem, each node exchanges messages with its neighbors according to a specific pattern. We define a new field for providing information on the number of RREQ messages. In fact, each node raises the number of its received messages by one (received++) after receiving a *RREQ* message and raises the number of its sent messages by one (sent++) after sending a *RREQ* message. Using this field guarantees the periodic tracing of the nodes behavior to check whether they are part of a flooding attack or not. Every node must contain the information of the *Hello* message about the exchanged *RREQ* packets. The *Hello* message is used not only to stabilize the connections between the neighbor nodes but also to check whether the network is secure regarding flooding attacks or not. Before sending the *Hello* message, nodes produce and receive some information about *RREQ* messages during sending the *Hello* message.

When a *Hello* message is received from a neighbor node, the receiving node marks its neighbor node as an active node and decodes the existing information in the *Hello* message. If the neighbor node is a new node, the receiving node creates a new input to record the information of this node in its table and writes the information of the node on but if it is a repetitive node, the receiving node updates the inputs related to. The node receiving the *Hello* message records every input in the table related to the neighbor which is selected as an active node and saves the information related to the exchanged *RREQ* packets. We assume that the unusual increased number of sent *RREQ* packets imply a flooding attack. We determine these unusual changes using the average transmission weight obtained from previous observations. If this weighted average exceeds the threshold, a detection process should be triggered by the attack detection system to detect the source of flooding attacks.

### 3.4 The mechanism to detect physical layer attack in LSFA-IoT

Detection of malicious nodes generated by fake RREQ packets in the IoT ecosystem is identified in the second step of the proposed method. In the IoT ecosystem, to discover a node that has generated a large number of RREQ packets, each node must search its neighbors list to find the node. Therefore, to detect Flooding attacks, each node calculates the number of RREQs generated using a weighted average formula in the proposed DAFA-DSR method.

APT-RREQ was used to calculate the average transfer of RREQ packets. In order to smooth the specified short-term and long-term fluctuations, the average of data transfer in a specified period of time has been used.

APT-RREQ for C series may be recursively calculated. In fact, to analyze and calculate the content of APT-RREQ, over a period of time, the number of RREQ packets observed has been used.

$$\begin{cases} D_1 = C_1 & \text{for } t = 1 \\ D_t = \alpha * C_t + (1-\alpha) * D_{t-1} & \text{for } t > 1 \end{cases} \quad (1)$$

As in Formula 1:

The coefficient α of the factor is a constant value between zero and one for smoothing.

$C_t$: Variable C is defined as how many RREQ packets are sent over a period of time. Some APT-RREQ is stored in $D_t$ in each time period t.

We use different values of α to intercept Flooding attacks in the proposed method. When the IoT network is subject to flooding attacks, low values of α are used to check for APT-RREQ. At the same time, high values of α can help analyze general network observations over a period of time and identify the source of the attack. Using the HELLO message, the number of RREQs transmitted for each node is specified. Each node calculates the neighboring node information by receiving the RREQ message. A node is detected as an attacker when the value assigned to the APTRREQ variable is greater than the threshold value defined. For this purpose, if the number of RREQ control messages sent by a node to neighboring nodes exceeds the threshold, that node is recognized as an Attacker.

### 3.5 Adding the malicious thing to detention list

When a thing detects a malicious neighbor thing, it adds that thing to its detention list and rejects all the requests received from that thing for a period of $\theta$. Also, it sends a $RREQ$ to its neighbors to eliminate their connections with the malicious thing and isolate that from the network for a period of $\theta$.

### 3.6 Revision of malicious thing

Each thing maintains the detention list field for $\theta = 4 * RTT$ where $RTT$ is the average round trip time of $RREQ$. When this deadline expires, the destructive node will change to the normal node and will participate in the routing process. When a node in the IoT ecosystem is identified as a healthy node, all neighboring nodes will update the proposed LSFA-IoT architecture input. If a duplicate node shows malicious behavior again, it will be placed in detention list again and all neighbor nodes make the changes in LSFA-IoT based on. As observed, we introduced our proposed LSFA-IoT method which is a useful method to detect flooding attacks in IoT. The flowchart of proposed LSFA-IoT is given in Fig. 4.

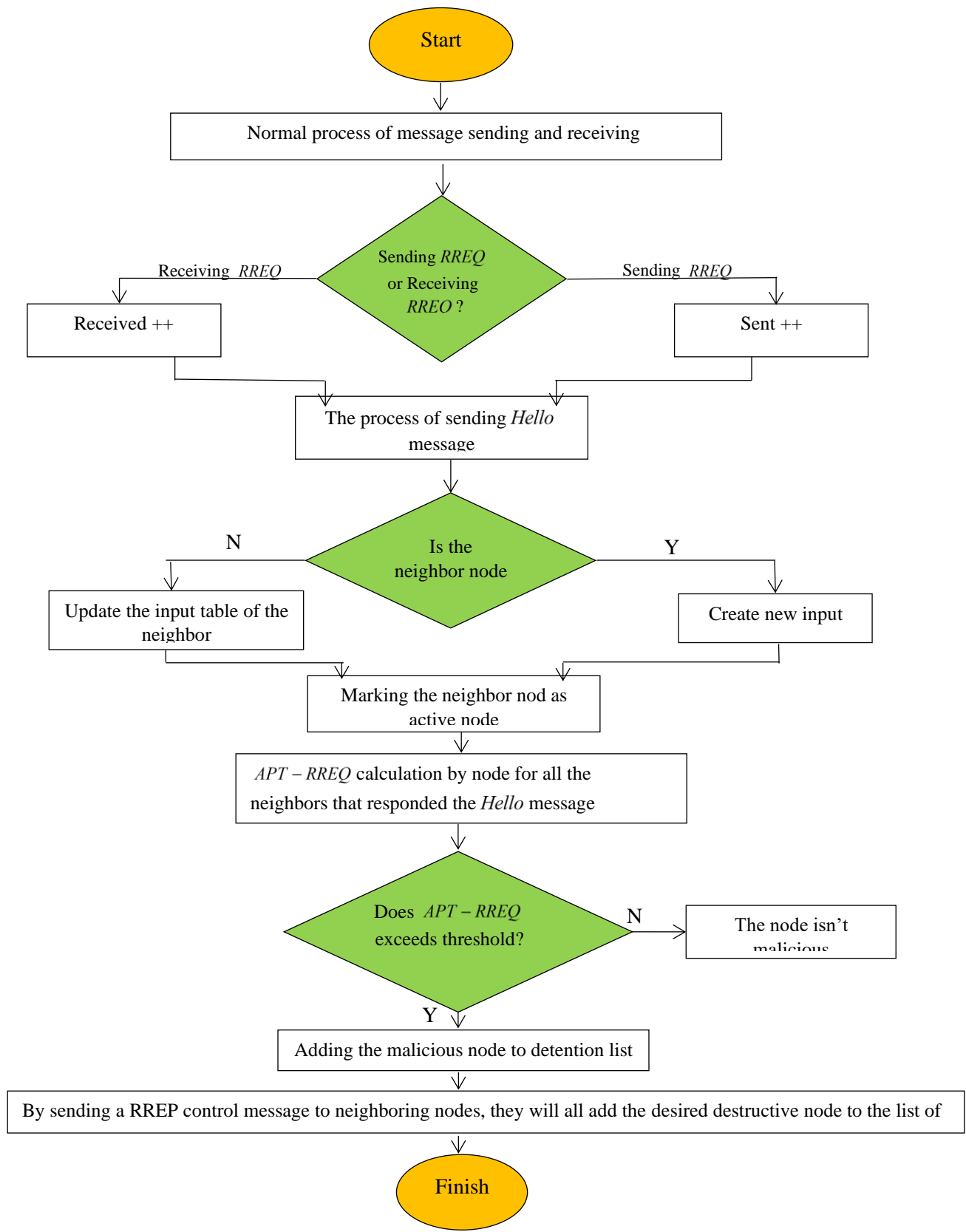

**Fig. 4** LSFA-IoT flowchart.

# 4 Evaluating the Performance

This section has two subsections called Performance metrics and Results: The results extracted from the simulator are shown in the format of tables and graphs. It also compares the simulation results of the proposed LSFA-IoT method with the last three methods that worked on the detection of malicious nodes under important criteria. The results were compared with two methods (REATO [16] and IRAD [15]). To demonstrate a feasibility study, the performance analysis of REATO, IRAD, and LSFA-IoT has been divided into five parts: FPR, FNR, DR, and Packet Delivery Rate (PDR).

## 4.1 Performance metrics

In this subsection, the concept and formula of all measure are explained in detail. The symbols are shown in Table 2.

**Table 2** Symbols.

| Symbols | Value |
|---|---|
| $A_i$ | Received packets by node i |
| $Z_i$ | Sent packets by node i |
| $Ex$ | Experiments |

**FPR:** A normal node has been identified as a malicious node [18-20]. Eq. (2) determines the FPR.

$$FPR = \left(\frac{FPR}{FPR+TNR}\right)*100 \quad \text{Where:} \quad TNR = \left(\frac{TNR}{TNR+FPR}\right)*100 \tag{2}$$

**FNR:** The malicious node has been identified as a normal node [21-23]. Eq. (3) determines the FNR.

$$FNR = \left(\frac{TPR+TNR}{All}\right)*100 \quad \text{Where:} \quad TPR = \left(\frac{TPR}{TPR+FNR}\right)*100 \tag{3}$$

**DR:** Total malicious nodes that were successfully detected [24-28]. Eq. (4) determines the DR.

$$DR = \left(\frac{TPR}{TPR+FNR}\right)*100 \quad \text{where} \quad All = TPR+TNR+FPR+FNR \tag{4}$$

**PDR:** Number of packets generated from the source node and successfully received at the destination node. This criterion is shown as a percentage. [29-35]. Eq. (5) determines the PDR.

$$PDR = \left(\frac{1}{Ex}*\frac{\sum_{i=1}^{n}A_i}{\sum_{i=1}^{n}Z_i}*100\%\right) \tag{5}$$

## 4.2 Results

In this subsection, simulation results are shown for all measures. The proposed LSFA-IoT method has been simulated and its performance evaluated in NS-3 Simulator on Linux Ubuntu 14 LTS. Because the data extracted from the simulation is correct and logical, the most important parameters of Things such as MAC Layer, etc. are used. The rest of the parameters used in the simulator are listed in Table 3. Details of the parameters used in the three scenarios are given in Table 4. The only difference between the three scenarios is the attacker rate. The rest of the parameters are considered the same.

**Table 3** Parameters used.

| Parameters | Value |
| --- | --- |
| Simulation tool | NS-3 |
| MAC | IEEE 802.11 |
| Transport | UDP/IPv6 |
| Range of communication | 300 m |
| Bandwidth | 3 Mbps |
| Traffic type, rate | CBR, 10 packets/sec |
| Model of mobility | Random way point |
| RX and TX ratio | 90% |
| Percentage of malicious nodes | 0% - 30% |
| Simulation time (varying) | 500-2000 |

**Table 4** Important parameters of all three scenarios.

| Scenario #1 | | Scenario #2 | |
| --- | --- | --- | --- |
| Attacks | 10% | Attacks | 20% |
| Topology (m x m) | 1000*1000 | Topology (m x m) | 1000*1000 |
| Time | 2000 | Time | 2000 |
| **Scenario #3** | | | |
| Attacks | 30% | | |
| Topology (m x m) | 1000*1000 | | |
| Time | 2000 | | |

Table 5 shows the average efficiency of all methods in a percentage of flooding attacks.

**Table 5** Comparison of all methods.

| Schemes | | Detection rate | FNR | FPR | PDR |
|---|---|---|---|---|---|
| IRAD | Number of IoT (10% of overall nodes) | 63.86 | 31.058 | 21.14 | 72.916 |
| | Number of IoT (20% of overall nodes) | 64.83 | 26.394 | 26.97 | 65.106 |
| | Number of IoT (30% of overall nodes) | 57.134 | 22.232 | 29.77 | 60.299 |
| REATO | Number of IoT (10% of overall nodes) | 70.84 | 27.23 | 19.04 | 77.837 |
| | Number of IoT (20% of overall nodes) | 67.458 | 21.749 | 22.232 | 69.54 |
| | Number of IoT (30% of overall nodes) | 65.91 | 18.43 | 26.392 | 67.227 |
| LSFA – IoT | Number of IoT (10% of overall nodes) | 88.819 | 14.583 | 12.011 | 93.28 |
| | Number of IoT (20% of overall nodes) | 87.31 | 11.629 | 16.13 | 83.394 |
| | Number of IoT (30% of overall nodes) | 83.005 | 10.305 | 17.68 | 76.947 |

FPR: The LSFA-IoT method has a lower FPR than the other approaches, as indicated in Table 6 and Figure 5. This is in a situation where the FPR of the LSFA-IoT method is lower than the methods compared in this paper. The reason for the superiority of the proposed method is the use of the APT-RREQ function to control the threshold of RREQ control packets, which if exceeded this threshold, the node is detected as malicious and is excluded from the routing process. The FPR of all methods is 15, 23, and 27% for the LSFA-IoT, REATO, and IRAD methods, respectively, when a malicious things rate is equal to 10%. The output for the second scenario is 18, 23, and 32, respectively, and it is equal to 20, 30, and 34, respectively, for the third scenario.

**Table 6** FPR (in %) of various frameworks with varying degree of malicious nodes.

| Misbehaving thing ratio | FPR (%) | | |
|---|---|---|---|
| | IRAD | REATO | LSFA – IoT |
| 0 | 8.3 | 9.25 | 2.42 |
| 0.05 | 10.31 | 12.24 | 3.65 |
| 0.10 | 19.05 | 19.01 | 5.41 |
| 0.15 | 27 | 23.35 | 6.57 |
| 0.20 | 34.38 | 28.62 | 9.63 |
| 0.25 | 47.6 | 31.88 | 11.97 |
| 0.30 | 52.67 | 39.09 | 13.83 |

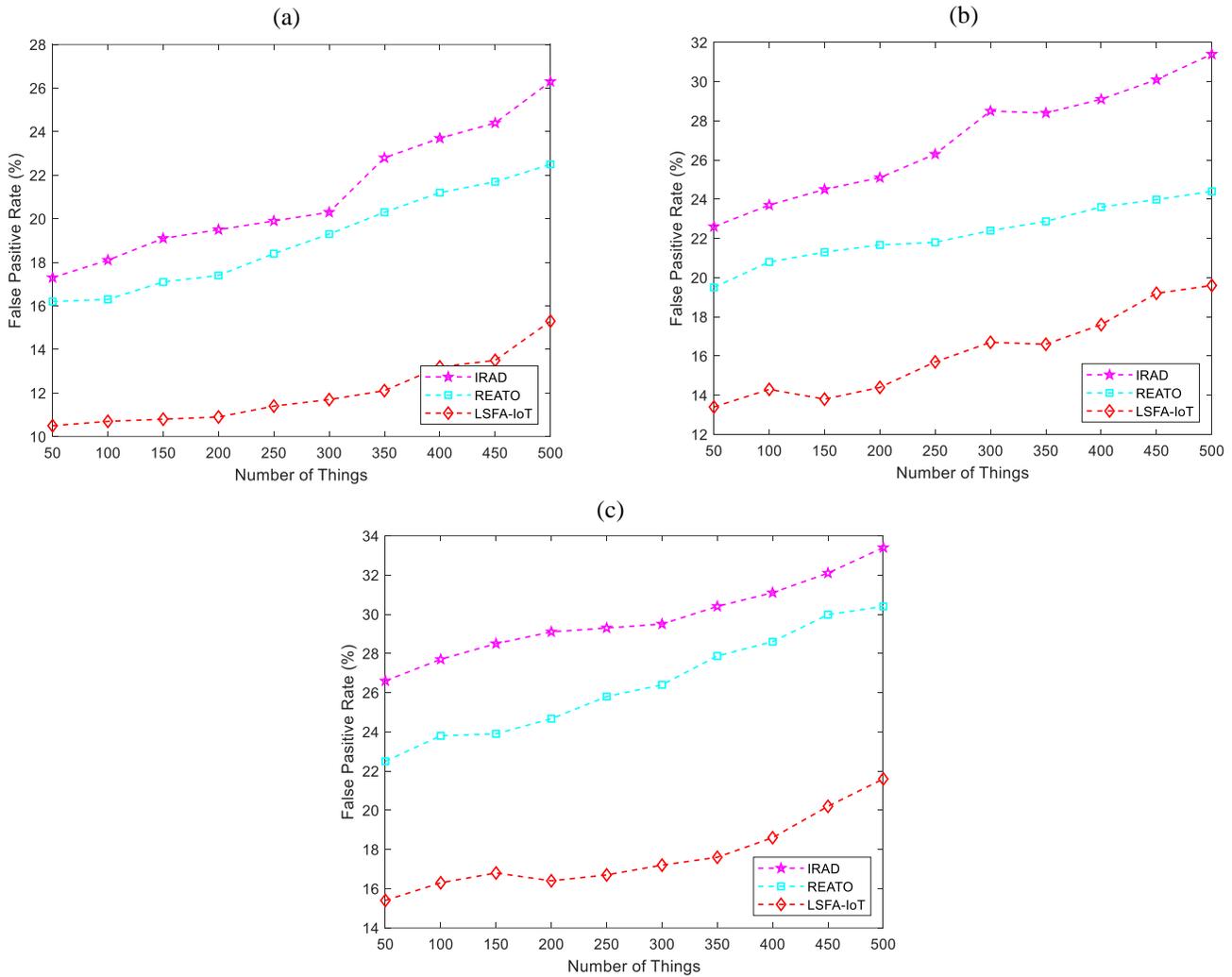

**Fig. 5** Comparison of FPR of all methods.

**FNR:** As you can see in Table 7 and Figure 6, the FNR method proposed in all three scenarios is better than the other two methods. When the range of enemy things increases from 250 to 500, the FNR criterion has grown less in the proposed method, but in all two methods, this growth has been very high, indicating that the enemy thing has had a greater impact on these two methods. The reason for the high FNR percentage in the other two methods, REATO, and IRAD is that they used only classical, rule-based methods to detect the position of the things, which has no effect on the stronger and safer identification of enemy things. The reason for the superiority of the proposed method is that the proposed LSFA-IoT performs a complete six-step investigation to identify flooding malicious nodes, and conversely, the decision is not based on a single factor. The FPR of all methods is 17, 32, and 37% for the LSFA-IoT, REATO, and IRAD methods, respectively, when a malicious things rate is equal to 10%. The output for the second scenario is 12, 25, and 31, respectively, and it is equal to 11, 20, and 25, respectively, for the third scenario.

**Table 7** FNR (in %) of various frameworks with varying degree of malicious nodes.

| Misbehaving thing ratio | FNR (%) | | |
|---|---|---|---|
| | *IRAD* | *REATO* | *LSFA − IoT* |
| 0 | 7.93 | 9.005 | 2.34 |
| 0.05 | 8.43 | 10.08 | 2.86 |
| 0.10 | 10.19 | 11.3 | 3.27 |
| 0.15 | 15.63 | 13.37 | 4.62 |
| 0.20 | 24.38 | 16.25 | 6.2 |
| 0.25 | 33.2 | 18.76 | 9.83 |
| 0.30 | 39.27 | 24.89 | 11.22 |

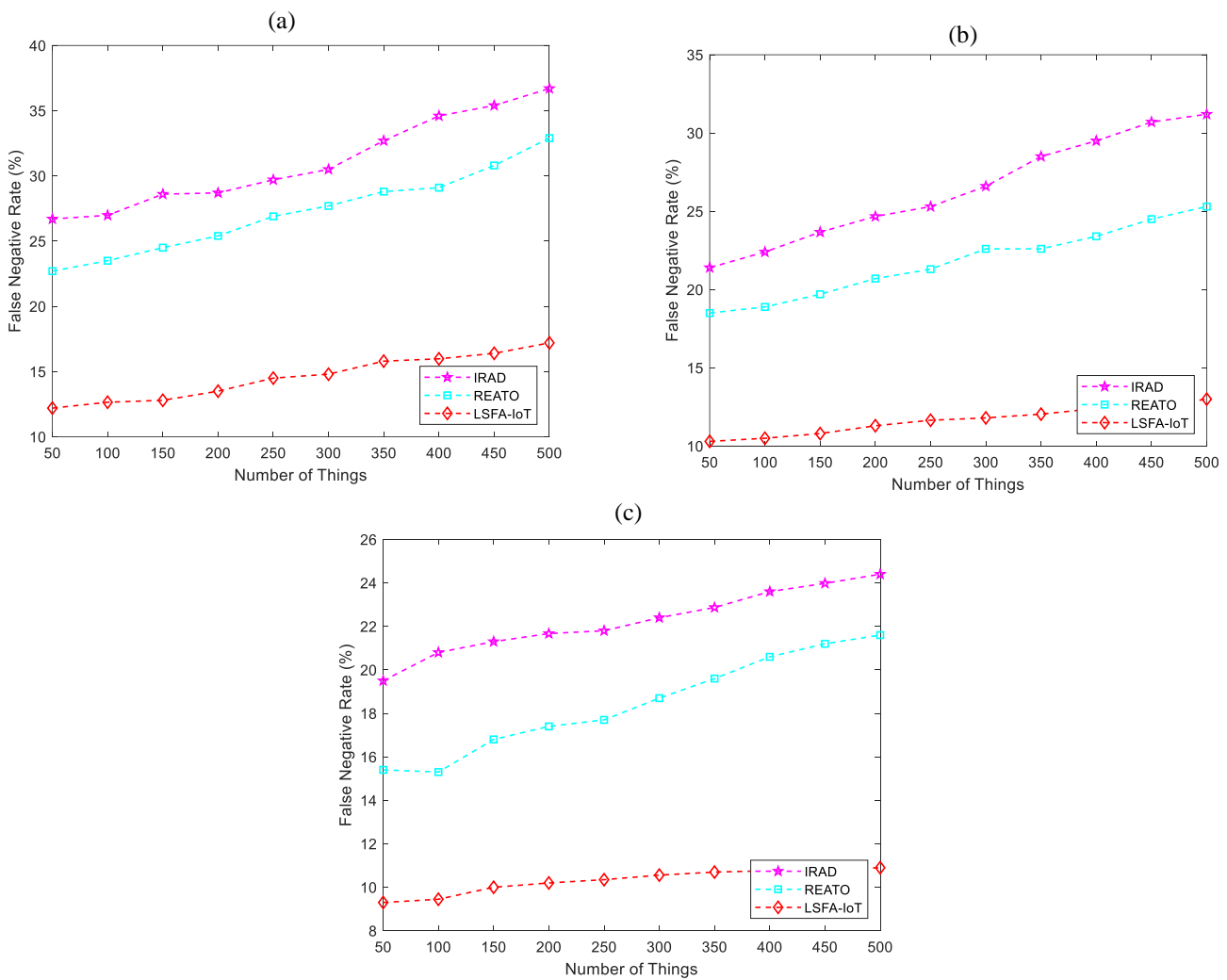

**Fig. 6** Comparison of FNR of all methods

**DR:** As you can see in Table 8 and Figure 7, the DR method proposed in all three scenarios is better than the other two methods. The reason for better DR in the proposed method is that the routes between the source and the destination must be discovered in the first step in order to choose the safest route among them. Therefore, in the first step, the route request packet (RREQ) is initially sent from the source Thing to the destination. However, in the proposed method, each thing maintains the detention list field for θ=4*RTT where RTT is the average round trip time of RREQ. When this deadline expires, the destructive node will change to the normal node and will participate in the routing process. When a node in the IoT ecosystem is identified as a healthy node, all neighboring nodes will update the proposed LSFA-IoT architecture input. If a duplicate node shows malicious behavior again, it will be placed in the detention list again and all neighbor nodes make the changes in LSFA-IoT based on.

**Table 8** DR (in %) of various frameworks with varying degree of malicious nodes.

| Misbehaving thing ratio | DR (%) | | |
| --- | --- | --- | --- |
|  | *IRAD* | *REATO* | *LSFA – IoT* |
| 0 | 91.63 | 90.2 | 97.5 |
| 0.05 | 89.49 | 88.57 | 96.2 |
| 0.10 | 80.46 | 81.8 | 94.38 |
| 0.15 | 73.35 | 76.37 | 92.27 |
| 0.20 | 63.19 | 70.43 | 90.28 |
| 0.25 | 50.34 | 66.16 | 87.7 |
| 0.30 | 46.14 | 60.67 | 84.4 |

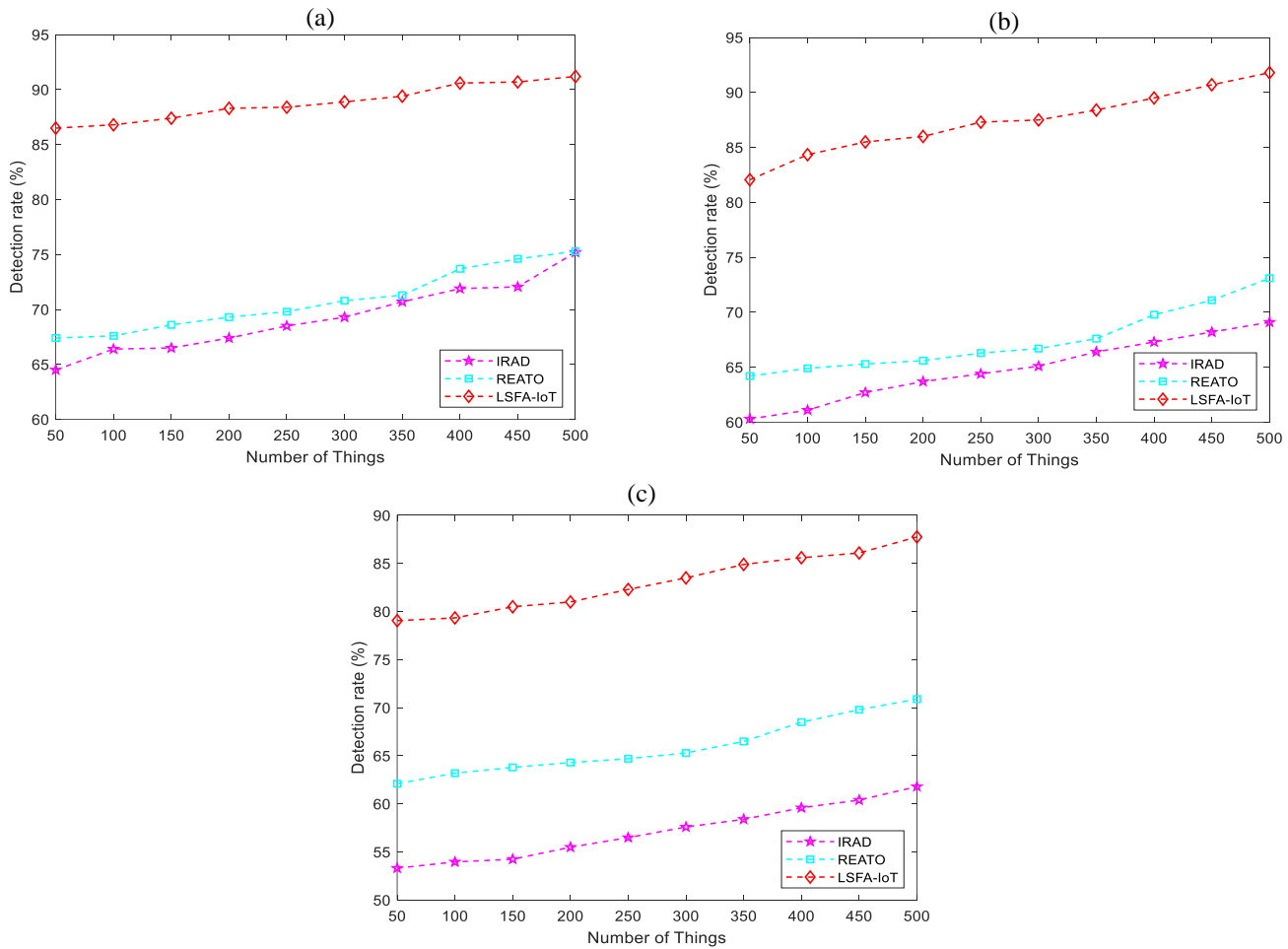

**Fig. 7** Comparison of the LSFA-IoT, REATO and IRAD.

PDR: Since the proposed approach selects the safest (most reliable) route from source to destination using an APT-RREQ, it has a better PDR than other methods. These results are represented in Table 9 and Figure 8. Therefore, while the malicious thing rate is 10%, the PDR of LSFA-IoT, REATO, and IRAD methods are 96%, 81%, and 75%, respectively. This output for the second scenario is 86, 75, and 63, respectively, and it is 81, 69, and 61, respectively, for the third scenario.

**Table 9** PDR (in %) of various frameworks with varying degree of malicious nodes.

| Misbehaving thing ratio | PDR (%) | | |
|---|---|---|---|
| | IRAD | REATO | LSFA – IoT |
| 0 | 84.4 | 86.4 | 98.7 |
| 0.05 | 76.1 | 80.1 | 95.4 |
| 0.10 | 70.1 | 73.1 | 93.1 |
| 0.15 | 62.3 | 68.3 | 90.2 |
| 0.20 | 55.13 | 60.13 | 87.1 |
| 0.25 | 49.2 | 53.2 | 84.3 |
| 0.30 | 35.23 | 45.23 | 81.2 |

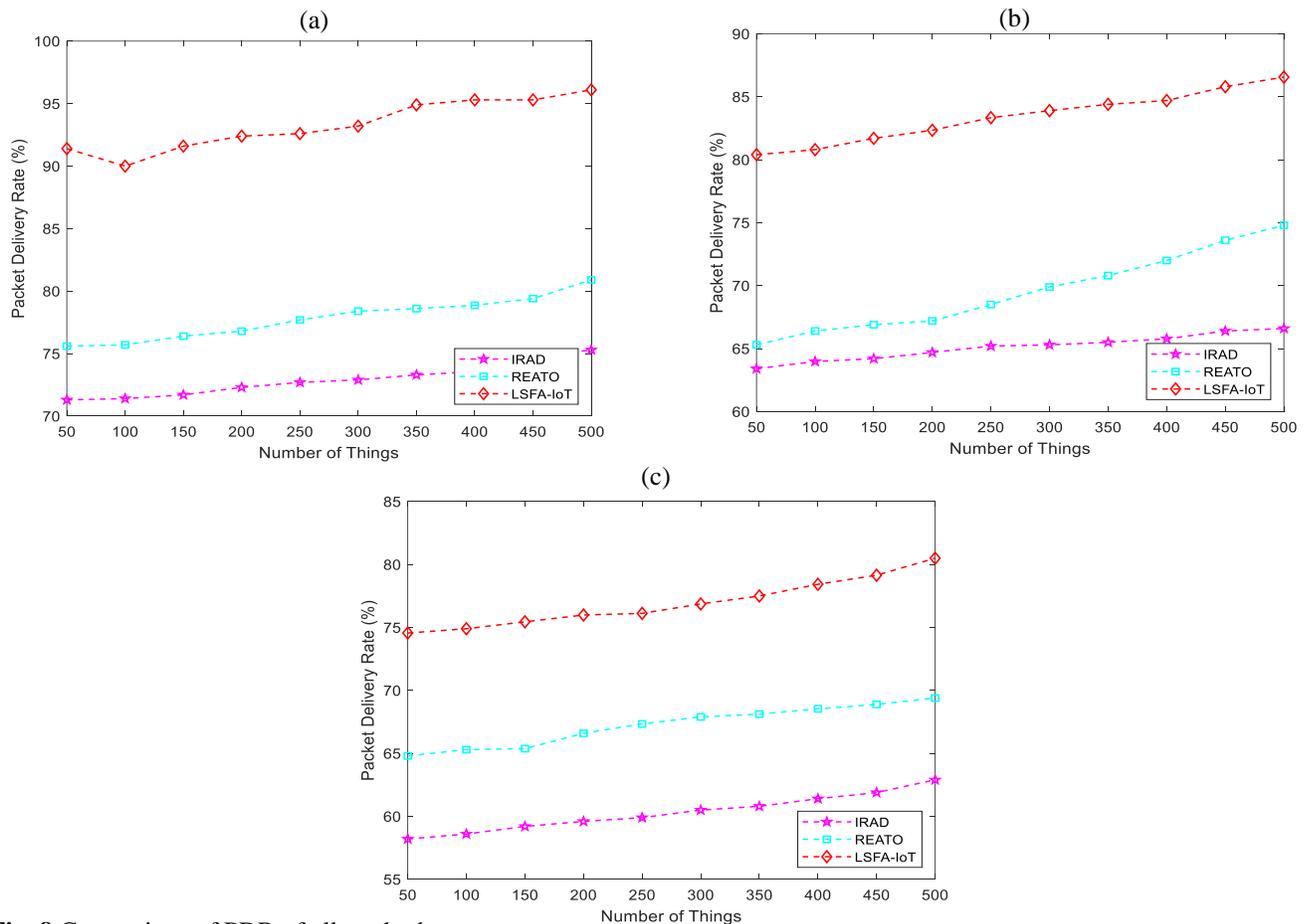

**Fig. 8** Comparison of PDR of all methods.

# 5 Conclusion

To overcome the insecurity in the IoT, and to achieve the secure state of the desired networks, a secure hybrid solution has been proposed. This solution consists of two basic phases: the first phase detects and identifies loading attacks, and in the second phase we eliminate the attacks discovered in the first phase so that they do not carry out malicious activity on the network. Also, the first section is utilized for stabilizing the network status. By exceeding the number of routes requests the threshold, the nodes become aware of abnormal behavior and misbehavior in the network. The second part is responsible for discovering the misbehavior sources in the network utilizing APT-RREQ, the Average Packet Transmission RREQ. By detecting a malicious node, the node's status is checked by LSFA-IoT prior to sending a data packet, and in case the node is found as a malicious node, the packet is not sent to that node and the node is added to detention list. The suggested LSFA-IoT can be utilized in effectively managing attacks within the route detection phase and over the data packet transmission phase. LSFA-IOT is more effective than the IRAD and REATO methods under flooding attack since it finds the malicious node previously in addition to isolating the malicious node and restoring the accused node followed by the penalty period. The proposed algorithm is found to effective on all the three different scenarios and attained optimal solutions in terms of FPR, FNR, DR, and PDR. In additional to, the simulation results show that the LSFA-IOT improves FPR, FNR, DR, and PDR significantly. This capability of LSFA-IOT provides the possibility of secure data transmission among IoT devices.

# Conflict of Interest

None.

# DATA Availability Statement

The data of this paper is the result of simulation and all the data are presented in the form of graphs inside the paper. There is no private data in this article.

# Funding

None

# Reference

1. Vishwakarma, R., & Jain, A. K. (2020). A survey of DDoS attacking techniques and defence mechanisms in the IoT network. Telecommunication Systems, 73(1), 3-25.
2. Furfaro, A., Pace, P., & Parise, A. (2020). Facing DDoS bandwidth flooding attacks. Simulation Modelling Practice and Theory, 98, 101984.
3. Tewari, A., & Gupta, B. B. (2020). Security, privacy and trust of different layers in Internet-of-Things (IoTs) framework. Future generation computer systems, 108, 909-920.
4. Stergiou, C., Psannis, K. E., Gupta, B. B., & Ishibashi, Y. (2018). Security, privacy & efficiency of sustainable cloud computing for big data & IoT. Sustainable Computing: Informatics and Systems, 19, 174-184.

5. Din, S., Paul, A., Ahmad, A., Gupta, B. B., & Rho, S. (2018). Service orchestration of optimizing continuous features in industrial surveillance using big data based fog-enabled internet of things. IEEE Access, 6, 21582-21591.
6. Tewari, A., & Gupta, B. B. (2017). Cryptanalysis of a novel ultra-lightweight mutual authentication protocol for IoT devices using RFID tags. The Journal of Supercomputing, 73(3), 1085-1102.
7. Stergiou, C. L., Psannis, K. E., & Gupta, B. B. (2020). IoT-based Big Data secure management in the Fog over a 6G Wireless Network. IEEE Internet of Things Journal.
8. Tabassum, A., Sadaf, S., Sinha, D., & Das, A. K. (2020). Secure Anti-Void Energy-Efficient Routing (SAVEER) Protocol for WSN-Based IoT Network. In Advances in Computational Intelligence (pp. 129-142). Springer, Singapore.
9. Zarpelao, B.B., et al., A survey of intrusion detection in Internet of Things. Journal of Network and Computer Applications, 2017. 84: p. 25-37.
10. Faghihniya, M.J., S.M. Hosseini, and M. Tahmasebi, Security upgrade against RREQ flooding attack by using balance index on vehicular ad hoc network. Wireless Networks, 2017. 23(6): p. 1863-1874.
11. Chen, K., et al., Internet-of-Things Security and Vulnerabilities: Taxonomy, Challenges, and Practice. Journal of Hardware and Systems Security, 2018. 2(2): p. 97-110.
12. Moon, J., I.Y. Jung, and J.H. Park, Iot application protection against power analysis attack. Computers & Electrical Engineering, 2018. 67: p. 566-578.
13. Jiang, Y., A. Hu, and J. Huang, A lightweight physical-layer based security strategy for Internet of things. Cluster Computing, 2018: p. 1-13.
14. Adat, V. and B. Gupta, Security in Internet of Things: issues, challenges, taxonomy, and architecture. Telecommunication Systems, 2018. 67(3): p. 423-441.
15. Yavuz, F. Y., Devrim, Ü. N. A. L., & Ensar, G. Ü. L. (2018). Deep learning for detection of routing attacks in the internet of things. International Journal of Computational Intelligence Systems, 12(1), 39-58.
16. Sicari, S., Rizzardi, A., Miorandi, D., & Coen-Porisini, A. (2018). REATO: REActing TO Denial of Service attacks in the Internet of Things. Computer Networks, 137, 37-48.
17. Alamr, A.A., et al., A secure ECC-based RFID mutual authentication protocol for internet of things. The Journal of Supercomputing, 2018. 74(9): p. 4281-4294.
18. Deng, L., et al., Mobile network intrusion detection for IoT system based on transfer learning algorithm. Cluster Computing, 2018: p. 1-16.
19. Bhunia, S.S. and M. Gurusamy. Dynamic attack detection and mitigation in IoT using SDN. in 2017 27th International Telecommunication Networks and Applications Conference (ITNAC). 2017. IEEE.
20. Rostampour, S., et al., A scalable and lightweight grouping proof protocol for internet of things applications. The Journal of Supercomputing, 2018. 74(1): p. 71-86.
21. Lee, S., et al., Design and implementation of cybersecurity testbed for industrial IoT systems. The Journal of Supercomputing, 2017: p. 1-15.
22. Hashemi, S.Y. and F.S. Aliee, Dynamic and comprehensive trust model for IoT and its integration into RPL. The Journal of Supercomputing, 2018: p. 1-30.
23. Diro, A.A. and N. Chilamkurti, Distributed attack detection scheme using deep learning approach for Internet of Things. Future Generation Computer Systems, 2018. 82: p. 761-768.
24. Jhaveri, R.H., et al., Sensitivity analysis of an attack-pattern discovery based trusted routing scheme for mobile ad-hoc networks in industrial IoT. IEEE Access, 2018. 6: p. 20085-20103.
25. Yaseen, Q., et al., Collusion attacks mitigation in internet of things: a fog based model. Multimedia Tools and Applications, 2018. 77(14): p. 18249-18268.
26. Bawany, N.Z., J.A. Shamsi, and K. Salah, DDoS attack detection and mitigation using SDN: methods, practices, and solutions. Arabian Journal for Science and Engineering, 2017. 42(2): p. 425-441.

27. Zaminkar, M., Sarkohaki, F., & Fotohi, R. (2021). A method based on encryption and node rating for securing the RPL protocol communications in the IoT ecosystem. International Journal of Communication Systems, 34(3), e4693.
28. Faraji-Biregani, M., & Fotohi, R. (2020). Secure communication between UAVs using a method based on smart agents in unmanned aerial vehicles. The Journal of Supercomputing, 1-28.
29. Rizvi, S., Pipetti, R., McIntyre, N., Todd, J., & Williams, I. (2020). Threat model for securing internet of things (IoT) network at device-level. Internet of Things, 11, 100240.
30. Fotohi, R., Nazemi, E., & Aliee, F. S. (2020). An Agent-Based Self-Protective Method to Secure Communication between UAVs in Unmanned Aerial Vehicle Networks. Vehicular Communications, 100267.
31. Zaminkar, M., & Fotohi, R. (2020). SoS-RPL: Securing Internet of Things Against Sinkhole Attack Using RPL Protocol-Based Node Rating and Ranking Mechanism. WIRELESS PERSONAL COMMUNICATIONS.
32. Fotohi, R., Firoozi Bari, S., & Yusefi, M. (2020). Securing wireless sensor networks against denial-of-sleep attacks using RSA cryptography algorithm and interlock protocol. International Journal of Communication Systems, 33(4), e4234.
33. Tukur, Y. M., Thakker, D., & Awan, I. U. (2020). Edge-based blockchain enabled anomaly detection for insider attack prevention in Internet of Things. Transactions on Emerging Telecommunications Technologies, e4158.
34. Jamali, S., & Fotohi, R. (2017). DAWA: Defending against wormhole attack in MANETs by using fuzzy logic and artificial immune system. the Journal of Supercomputing, 73(12), 5173-5196.
35. Butun, I., Österberg, P., & Song, H. (2019). Security of the Internet of Things: Vulnerabilities, attacks, and countermeasures. IEEE Communications Surveys & Tutorials, 22(1), 616-644.